\documentclass[showpacs,amsmath,amssymb,twocolumn,nofootinbib]{revtex4}
\usepackage{epsfig}
\input{epsf}
\usepackage{amssymb}
\begin{document}
\title{ The $\mbox{\boldmath$a$}$-theorem and temperature of the CMB temperature in cosmology}
\author{A.O.Barvinsky}
\affiliation{Theory Department, Lebedev Physics Institute, Leninsky
Prospect 53, 119991 Moscow, Russia}






\begin{abstract}
Initial conditions in cosmology in the form of the microcanonical density matrix of the Universe predict a thermal nature of the primordial CMB power spectrum with a nonzero temperature of the resulting relict temperature distribution. This effect generates a thermal contribution to the red tilt of this spectrum, additional to its vacuum component. In the cosmological model with a large number of free fields conformally coupled to gravity the magnitude of this effect is determined by the Gauss-Bonnet coefficient $\mbox{\boldmath$a$}$ of the trace anomaly. For low spins it is too small to be presently observable, but it can be amplified by the mechanism of the $\mbox{\boldmath$a$}$-theorem applied to the renormalization group flow which interpolates between the ultraviolet and infrared domains associated respectively with early and late stages of cosmological evolution.
\end{abstract}

\pacs{04.60.Gw, 04.62.+v, 98.80.Bp, 98.80.Qc}
\maketitle

\section{Introduction}
As has recently been persuasively advocated, the renormalization group (RG) flow in 4D conformal field theory (CFT) is subject to the so-called $\mbox{\boldmath$a$}$-theorem \cite{KomargodskiSchwimmer,Komargodski} -- the analogue of the 2D $c$-theorem \cite{Zamolodchikov}. For the 4D CFT embedded in curved spacetime with metric $g_{\mu\nu}$ and having the effective action $W[\,g_{\mu\nu}]$,
    \begin{eqnarray}
    e^{iW[\,g_{\mu\nu}]}=\int D\phi\,e^{iS_{CFT}[\,g_{\mu\nu},\phi\,]}, \label{W0}
    \end{eqnarray}
this is a statement, made on the basis of the trace anomaly matching \cite{matching}, that the coefficient $\mbox{\boldmath$a$}$ of the Gauss-Bonnet invariant $E=R_{\mu\nu\alpha\gamma}^2- 4R_{\mu\nu}^2+ R^2$ in the trace anomaly
    \begin{equation}
    \langle\, T^\mu_\mu\,\rangle\equiv
    \frac2{g^{1/2}}\,g^{\mu\nu}\frac{\delta
    W}{\delta g^{\mu\nu}} =
    \mbox{\boldmath$a$} E -\mbox{\boldmath$c$}\, C_{\mu\nu\alpha\beta}^2
    +\mbox{\boldmath$b$}\, \Box R         \label{Wanomaly}
    \end{equation}
monotonically decreases in the course of this flow from the ultraviolet (UV) to infrared (IR) limits of the theory. According to \cite{KomargodskiSchwimmer,Komargodski} the difference between the UV and IR values of $\mbox{\boldmath$a$}$ is related to the total cross section $\sigma(s)=s\,{\rm Im} {\cal A}(s,t)_{t=0}>0$ of the forward $2\to2$ scattering of the dilaton -- Nambu-Goldstone boson of broken conformal symmetry, and the positivity of $\sigma(s)$ in unitary theory provides the positive increment of running $\mbox{\boldmath$a$}(\mu^2)$
    \begin{eqnarray}
    \mbox{\boldmath$a$}_{UV}
    -\mbox{\boldmath$a$}_{IR}=
    \frac1{4\pi}\int_{s>0} ds\,
    \frac{\sigma(s)}{s^2}>0.        \label{dispersionrelation}
    \end{eqnarray}

The purpose of this Letter is to suggest a possible application of this theorem  -- enhancement of the red tilt of the CMB spectrum in the CFT driven cosmology due to the thermal (rather than vacuum \cite{ChibisovMukhanov}) origin of this spectrum from the microcanonical initial conditions in cosmology \cite{slih,why}. These initial conditions were recently put forward as the concept of the density matrix of the Universe -- the projector onto the subspace of physical states satisfying the full system of the Wheeler-DeWitt equations \cite{why}. When applied to the CFT cosmology this concept suggests a number of interesting conclusions including a limited range of the effective cosmological constant in the early Universe \cite{slih} and prediction of a thermal primordial CMB spectrum -- a nonvanishing temperature of the CMB temperature in cosmology \cite{bigboost}. Here we show that due to the $\mbox{\boldmath$a$}$-theorem amplification of this effect is possible within the RG flow, which interpolates between the early and present states of the Universe, by increasing in the UV the number of higher-spin conformal degrees of freedom.

\section{Dilaton dynamics}
Critical point associated with the $\mbox{\boldmath$a$}$-theorem is that the Gauss-Bonnet invariant itself, being a total derivative part of UV divergences, never contributes to local dynamics of the theory and seemingly does not lead to any positivity bounds. Indeed, in contrast to $E$-invariant, the Weyl squared part of the effective action contributes not only to UV divergences, which are just the integrated conformal anomaly (\ref{Wanomaly}) (say, in dimensional regularization with $d\to 4$), but also to their finite tail -- the logarithmic nonlocal part
    \begin{eqnarray}
    &&\!\!\!\!\!\!\!\!\!\!W=\frac1{32\pi^2(2-\frac{d}2)}
    \int d^4x\, g^{1/2}\Big(\mbox{\boldmath$c$}\,C_{\mu\nu\alpha\beta}^2
    -\mbox{\boldmath$a$}\,E\Big)\nonumber\\
    &&
    -\frac{\mbox{\boldmath$c$}}{32\pi^2}\int d^4x\,g^{1/2} C_{\mu\nu\alpha\beta}
    \ln\frac{-\Box
    -i\varepsilon}{\mu^2}\,C^{\mu\nu\alpha\beta}.                        \label{W}
    \end{eqnarray}
In the momentum representation the nonlocal logarithm equals
    $\ln[(-\Box
    -i\varepsilon)/\mu^2]=
    \ln(|p^2|/\mu^2)-i\pi\theta(-p^2)$,
so that the imaginary part of the effective action is quadratic in the Fourier transform of the Weyl tensor $\hat C_{\mu\nu\alpha\beta}(p)$
    \begin{eqnarray}
    {\rm Im} W=\frac{\mbox{\boldmath$c$}}{32\pi}\int d^4p\, \hat C_{\mu\nu\alpha\beta}^2(p)\,\theta(-p^2).
    \end{eqnarray}
Then, unitarity, ${\rm Im} W>0$, implies positivity of $\mbox{\boldmath$c$}$. No such restriction holds for $\mbox{\boldmath$a$}$, because the $E$-term in the divergent part of the action (\ref{W}) being a total derivative does not have a logarithmic counterpart among finite nonlocal terms of $W$. Nevertheless, the Gauss-Bonnet coefficient $\mbox{\boldmath$a$}$ is not dynamically inert but rather effects the scattering of the dilaton field -- the parameter of broken local Weyl invariance. As shown in \cite{KomargodskiSchwimmer,Komargodski}, unitarity of this scattering gives a restriction on the RG flow of $\mbox{\boldmath$a$}$, which is more complicated than a simple restriction on the sign of $\mbox{\boldmath$c$}$. This dilaton field and its action induced by $E$-part of the trace anomaly both arise as a consequence of this symmetry breakdown and can be derived by the Wess-Zumino procedure of anomaly integration along the orbit of the conformal group $g_{\mu\nu}=e^\sigma \bar g_{\mu\nu}$,
    \begin{equation}
    \frac{\delta \varGamma[\,e^\sigma\bar g\,]}{\delta\sigma}\,=g^{1/2}
    \big(\mbox{\boldmath$b$}\, \Box R +
    \mbox{\boldmath$a$}\, E -\mbox{\boldmath$c$}\, C_{\mu\nu\alpha\beta}^2\big)
    \Big|_{\;g\,=\,e^\sigma\bar g}\,.        \label{orbiteq}
    \end{equation}
The resulting Wess-Zumino action for $\sigma$ is just the difference of effective actions calculated on two members of this orbit $g_{\mu\nu}$ and $\bar g_{\mu\nu}$. It reads \cite{anomalyaction}
    \begin{eqnarray}
    &&\!\!\!\!\!\!\!\!\!\!\!\!\!\!\varGamma[\,g\,]-\varGamma[\,\bar g\,]=
    \frac12\,\int d^4x \bar g^{1/2} \left\{\vphantom{\frac23}\,\mbox{\boldmath$a$}\,\sigma{\cal \bar D}\sigma
    \right.\nonumber\\
    &&\left.+\Big[\,\mbox{\boldmath$a$}\,\Big(\bar E-\frac{2}{3}\bar\Box \bar R\Big)-\mbox{\boldmath$c$}\, \bar C_{\mu\nu\alpha\beta}^2\,\Big]\,\sigma
    \right\}\nonumber\\
    &&-\Big(\,\frac{\mbox{\boldmath$a$}}{18}
    +\frac{\mbox{\boldmath$b$}}{12}\,\Big)
    \int d^4x\,\Big(g^{1/2}R^2-
    \bar{g}^{1/2}\bar R^2\Big),              \label{RTF}
    \end{eqnarray}
where all barred quantities are built in terms of $\bar g_{\mu\nu}$ and ${\cal D}={\cal D} = \Box^2 + 2R^{\mu\nu}\nabla_{\mu}\nabla_{\nu} -
    \frac{2}{3} R\,\Box
    + \frac{1}{3}(\nabla^{\mu}R)\nabla_{\mu}$.
This operator has a number of special properties, including the local Weyl invariance of its densitized version
    $\bar g^{1/2}{\cal\bar D}=g^{1/2}{\cal D}$
and the linear transformation law for the Gauss-Bonnet density (modified by the $\Box R$ term)
    $g^{1/2}(E-\frac23\,\Box R)=
    \bar g^{1/2}(\bar E-\frac23\,\bar\Box\bar R)
    +2\,\bar g^{1/2}{\cal\bar D}\sigma$.                     
Absence of its local Weyl invariance -- ``non-abelian" nature of the Gauss-Bonnet anomaly -- is a main source of the nontrivial Wess-Zumino action \cite{KomargodskiSchwimmer}.

Integration of Eq.(\ref{orbiteq}) a priori leads to a fourth-order polynomial in $\sigma$, whose cubic and quartic terms are collected in (\ref{RTF}) into the curvature squared invariant, $\int d^4x\,\big(g^{1/2}R^2(g)-\bar g^{1/2} R^2(\bar g)\big)$. Therefore, the finite renormalization by the local counterterm (admissible from the viewpoint of UV renormalization), $\varGamma\to \varGamma_{R}
    =\varGamma
    +\frac1{12}\,\mbox{\boldmath$b$}\int d^4x\,g^{1/2}R^2$,
effectively puts $\mbox{\boldmath$b$}$ to zero and yields the following {\em minimal} dilaton action
    \begin{eqnarray}
    &&\!\!\!\!\!\!\varGamma_R[\,g\,]-\varGamma_R[\,\bar g\,]=
    -\frac{\mbox{\boldmath$c$}}2\int d^4x\,\bar g^{1/2}\,\sigma
    \,\bar C_{\mu\nu\alpha\beta}^2             \nonumber\\
    &&\,\,
    +\,\mbox{\boldmath$a$}\int d^4x\,\bar g^{1/2}\! \left\{\frac{1}{2}\,
    \sigma\bar E-\Big(\bar R^{\mu\nu}-\frac12\bar g^{\mu\nu}\bar
    R\Big)\partial_\mu\sigma\,\partial_\nu\sigma \right.  \nonumber\\
    &&\qquad\quad\left.-
    \,\frac12\,\bar\Box\sigma\,
    (\bar\nabla^\mu\sigma\,\bar\nabla_\mu\sigma)
    -\frac18\,(\bar\nabla^\mu\sigma\,
    \bar\nabla_\mu\sigma)^2\right\}.         \label{minimal}
    \end{eqnarray}

This expression was used in \cite{KomargodskiSchwimmer} for the derivation of the $\mbox{\boldmath$a$}$-theorem (\ref{dispersionrelation}), unitarity (and, therefore, positivity of (\ref{dispersionrelation})) being guaranteed by the absence of higher-derivative ghosts in (\ref{minimal}). The dynamical nature of the dilaton and its contribution to $\mbox{\boldmath$a$}$ was, however, later retracted in \cite{Komargodski} where $\sigma$ was assumed to be an external field never forming quantum loops. In contrast to the usual CFT setup, using $g_{\mu\nu}$ merely as an auxiliary tool which probes stress tensor correlators, in cosmology $\sigma$ serves as a physical observable -- the metric scale factor. Still it is not dynamically independent, but for another reason -- due to gravitational constraint equations, and we will consider its dynamical properties in more detail.

To begin with, absence of hihger-derivative modes of $\sigma$ in (\ref{minimal}) is the result of renormalization of $\mbox{\boldmath$b$}$ to zero, because quartic derivatives of $\sigma$ completely cancel out in the combination $\sigma\bar{\cal D}\sigma-\frac19 e^{2\sigma}R^2(e^\sigma\bar g)$ of Eq.(\ref{RTF}) \cite{slih}. This is true also for the third term in curly brackets of Eq.(\ref{minimal}), because its variation yields maximum second order derivatives
    \begin{eqnarray}
    &&\!\!\!\!\!\!\!\!\!\!\!\!\!\!\frac{\delta}{\delta\sigma}
    \int d^4x \bar g^{1/2}\,\bar\Box\sigma\,
    (\bar\nabla^\mu\sigma\,\bar\nabla_\mu\sigma)\nonumber\\
    &&\!\!\!\!\!\!=
    2\,\bar g^{1/2}\Big((\bar\nabla_\mu\bar
    \nabla_\nu\sigma)^2
    -(\bar\Box\sigma)^2
    +\bar R^{\mu\nu}\bar\nabla_\mu
    \sigma\bar\nabla_\nu\sigma\Big)    \label{sigmaeq}
    \end{eqnarray}
and for $\nabla_\mu\sigma\neq 0$ forces the characteristic surface (sound cone) to deviate from the light cone,
    \begin{eqnarray}
    &&\!\!\!\!\!\!\!\!\!\!\!\!\!\!\!\!\!\frac1{4\bar g^{1/2}}\frac{\delta^2}{\delta\sigma(y)\,\delta\sigma(x)}
    \int d^4x \bar g^{1/2}\,\bar\Box\sigma\,
    (\bar\nabla^\mu\sigma\,\bar\nabla_\mu\sigma)\nonumber\\
    &&\!\!\!\!\!\!\!\!\!\!\!\!\!\!\!\!\!=
    \big[\big(\bar\nabla^\mu\bar\nabla^\nu\sigma-\bar g^{\mu\nu}\bar\Box\sigma\big)
    \bar\nabla_\mu\bar\nabla_\nu+\bar R^{\mu\nu}\bar\nabla_\mu\sigma
    \bar\nabla_\nu\big] \delta(x,y).   \label{kinetic}
    \end{eqnarray}
This cubic term plays a special role in recent modifications of gravity theory like brane induced gravity models and massive graviton models where it survives the so-called decoupling limit \cite{DGP,boxnablapi} and represents the ghost-free higher-derivative braiding of metric and matter \cite{Vikmanetal}.

\section{CFT driven cosmology}
The CFT cosmology is governed by the Einstein-Hilbert action with the cosmological constant $\varLambda$ and the matter action dominated by many field multiplets $\phi$ conformally coupled to metric \cite{slih}. Their microcanonical statistical sum has the representation of the {\em Euclidean} path integral over periodic metric and matter fields with their full action $I[g_{\mu\nu},\phi]$ \cite{why}. Integrating $\phi$ out,
    \begin{equation}
    Z=\int D[g_{\mu\nu},\phi]\,
    e^{-I[g_{\mu\nu},\phi]}=
    \int Dg_{\mu\nu}\,
    e^{-I_{\rm eff}[g_{\mu\nu}]},         \label{Z}
    \end{equation}
one obtains the effective action
    \begin{equation}
    I_{\rm eff}[g_{\mu\nu},\phi]=-\frac1{16\pi G}
    \int d^4x\,g^{1/2}(R-2\varLambda)
    +\varGamma[g_{\mu\nu},\phi],    \label{tree}
    \end{equation}
where $\varGamma[g_{\mu\nu}]=-iW[g_{\mu\nu}]$ is the Euclidean version of the CFT effective action (\ref{W0}).

Physics of the CFT driven cosmology is entirely determined by this effective action. Solutions of its equations of motion, which give a dominant contribution to the statistical sum, are the cosmological instantons of $S^1\times S^3$ topology, which have the Friedmann-Robertson-Walker metric
    $g_{\mu\nu}dx^\mu dx^\nu=N^2(\tau)\,d\tau^2
    +a^2(\tau)\,d^2\Omega^{(3)}$
with the periodic scale factor $a(\tau)$ -- the function of the Euclidean time belonging to the circle $S^1$ \cite{slih}. These instantons serve as initial conditions for the cosmological evolution $a_L(t)$ in the physical Lorentzian spacetime. The latter follows from $a(\tau)$ by analytic continuation $a_L(t)=a(\tau_++it)$ at the point of the maximum value of the Euclidean scale factor $a_+=a(\tau_+)$. As it was discussed in \cite{bigboost} this evolution can incorporate a finite inflationary stage if the model (\ref{tree}) is generalized to the case when $\varLambda$ is replaced by a composite operator $\varLambda(\phi)=8\pi GV(\phi)/3$ -- the potential of the inflaton $\phi$ staying in the slow-roll regime during the Euclidean and inflationary stages and decaying in the end of inflation by a usual exit scenario.

In cosmology the dilaton $\sigma$ corresponds to the scale factor $a$ in the FRW metric and the conformal mode of the metric perturbations. Like in Einstein theory, it is non-dynamical because it is eliminated by the Hamiltonian constraint, even though the variation of (\ref{minimal}) with respect to the lapse function $N$ contains second order derivative $\ddot\sigma$, because the latter expresses in terms of $\sigma$ and its spatial gradients from the variational equation for $\sigma$ -- its linearity in $\ddot\sigma$ (cf. Eq.(\ref{sigmaeq})) allows one to do it. This mechanism looks even simpler in the long wavelengths limit of the scalar sector of cosmological variables, which corresponds to the FRW metric and spatially homogeneous dilaton $\sigma(\tau)$.

In this case $\bar g_{\mu\nu}$ in (\ref{minimal}) should be
identified with the metric of the Einstein static Universe of a unit radius, $d\bar s^2=d\eta^2+d\Omega_3^2$, $\eta$ is a conformal time, $d\eta=d\tau\,N/a$, and $e^\sigma=a^2$. Then the action (\ref{minimal}) on the FRW background reads
    \begin{eqnarray}
    &&\varGamma_R[\,g\,]-\varGamma_R[\,\bar g\,]
    =\frac{3\beta}4\int_{S^1} d\tau N
    \left(\frac{a'^2}{a}
    -\frac{a'^4}{6 a}\right), \label{anomalyaction}\\
    &&\beta=32\pi^2 \mbox{\boldmath$a$}.             \label{beta}
    \end{eqnarray}
Here $a'=da/Nd\tau$, and we introduce a new notation $\beta$ for the rescaled $\mbox{\boldmath$a$}$-coefficient (to avoid confusion with the notation for the FRW scale factor and match with notations of \cite{slih,why}). It does not contain $a''$, because $\bar\Box\sigma(\bar\nabla\sigma)^2\sim
\sigma''(\sigma')^2$ is a total derivative, and $\varGamma_R[\,\bar g\,]$ consists of the free energy $F(\eta)$ and the contribution of the vacuum Casimir energy $E^{\rm vac}_R$, $\varGamma[\,\bar g\,]=F(\eta)+ E^{\rm vac}_R\eta$ \cite{slih}. Here $F(\eta)=\pm\sum_{\omega} \ln\big(1\mp e^{-\omega\eta}\big)$ is a typical boson or fermion sum over conformal field oscillators with energies $\omega$ on a unit 3-sphere with the temperature given by the inverse of the Euclidean time period in units of the conformal time $\eta=\oint_{S^1} d\tau N/a$, and $E^{\rm vac}_R=\sum_\omega\omega/2|_{\,\rm renorm} =3\beta/8$ is a covariantly renormalized Casimir energy \cite{slih}.

These expressions together with the Einstein-Hilbert term lead to the total effective action derived in \cite{slih}, $I_{\rm eff}[\,a,N\,]= \frac{3\pi}{4G}\int d\tau\,N {\cal L}(a,a')+ F(\eta)$, with the local Lagrangian
    \begin{equation}
    {\cal L}(a,a')=-aa'^2
    -a+ \frac{\varLambda}3a^3
    +\frac{\beta G}\pi\!\left(\frac{a'^2}{a}
    -\frac{a'^4}{6 a}+\frac1{2a}\right).       \label{Gamma}
    \end{equation}
By the variation of $N$ it gives the Hamiltonian constraint and indicates the absence of dynamical modes in the minisuperspace sector, mentioned above. This nonlocal Euclidean Friedmann equation for $a(\tau)$ generates periodic cosmological instantons which specify the initial state of the Universe and give rise to inflationary evolution. The latter is driven by the inflaton $\phi$ with $\varLambda=\varLambda(\phi)$, which in view of the non-dynamical nature of $a$ turns out to be a single dynamical degree of freedom in the scalar sector of the theory. This dynamics crucially depends on the overall Gauss-Bonnet coefficient (\ref{beta}). For the set of $\mathbb{N}_s$ free field multiplets of spin $s$ their contributions to $\beta$ are weighted by well-known coefficients $\beta_s$, $\beta=\sum_s\beta_s\mathbb{N}_s$.

Due to thermal nature of the initial state all modes, including non-conformal inflaton and cosmological perturbations, have a spectrum modified by the Boltzmann factor $1/(e^{k\eta}-1)$ (cf. the expression for bosonic $F(\eta)$). With the comoving momentum of the non-conformal modes denoted by $k$ (contrary to $\omega$ for conformal ones)  the power spectrum of cosmological perturbations reads
    \begin{eqnarray}
    &&\!\!\!\!\!\!\!\!\!\!\!\!\!\!\delta_\phi^2(k)=
    \langle\, \hat a^\dagger_k
    \hat a_k+\hat a_k
        \hat a^\dagger_k\,\rangle_{\rm thermal}
    \,|u_k(t)|^2\nonumber\\
    &&\,=|u_k(t)|^2\big(1+{2N_k(\eta)}\big),
    \quad N_k(\eta)=\frac1{e^{k\eta}-1},       \label{occupation}
    \end{eqnarray}
where $u_k(t)$ is the positive frequency basis function in the $k$-mode and $N_k(\eta)$ is the occupation number in the thermal state with the ``comoving" temperature $1/\eta$.
In particular, the spectral index acquires the thermal contribution,
$n_s(k)=n_s^{\rm vac}(k)+\Delta n_s^{\rm thermal}(k)$,
    \begin{eqnarray}
    \Delta n_s^{\rm thermal}(k)
    =\frac{d}{d\ln k}\ln\big(1+{2N_k(\eta)}\big).
    \end{eqnarray}

For high temperature instantons the comoving temperature
equals $1/\eta=(180 \tilde\beta)^{1/6}/2\pi\gg 1$  \cite{CMBA-theorem}, where $\tilde\beta$ is a specific value of $\beta$ per one conformal degree of freedom
    \begin{equation}
    \tilde\beta=\frac1{\mathbb{N}}\sum_s\beta_s \mathbb{N}_s, \quad\mathbb{N}=\mathbb{N}_0+2\left(\frac78\, \mathbb{N}_{1/2}+\mathbb{N}_1\right),
    \end{equation}
with $\mathbb{N}$ -- the effective number of degrees of freedom modified by the well-known coefficient 7/8 for the thermal contribution of fermions. Therefore, the Boltzmann factor with the comoving momentum expressed in terms of the CMB multipole number $l$, $k_l\eta=l\,a_0 H_0\eta/\pi=l\,(\Omega_0-1)^{-1/2}\eta/\pi$, reads $N_{k_l}\simeq
\exp[-2l/(\Omega_0-1)^{1/2}(180\tilde\beta)^{1/6}]$, where $\Omega_0\simeq 1.01$ is the present cosmological density parameter of the closed cosmology. Thus, 
    \begin{eqnarray}
    \Delta n_s^{\rm thermal}(k_l)\simeq-2\,k\eta\,e^{-k\eta}\simeq
    -\frac{20\,l}{(3\tilde\beta)^{1/6}}
    e^{-\frac{10\,l}{(3\tilde\beta)^{1/6}}},
    \end{eqnarray}
which is too small to be observed at the pivotal scale $l\sim 10$ for low-spin models with $3\tilde\beta=O(1)$.

Enhancing the thermal effect can be based on climbing up the ladder of higher spins with the increasing $\beta_s$ \cite{ChristensenDuff}, as shown for conformal invariant gravitino and graviton (Weyl theory) \cite{Tseytlinconf} as compared to low spins -- real scalar, Dirac spinor and vector multiplets,
    \begin{equation}
    \beta_s\!=\!\frac1{180}\!\times\!
    \left\{\begin{array}{cl}  1 & s=0\\
     11 & s=\frac12\\
     62 & s=1
    \end{array}\right.\!,\;\;
    \beta_s^{\rm Weyl}\!=\!\frac1{180}\!\times\!
    \left\{\begin{array}{cl} -548 & s=\frac32\\
    1566 & s=2\\
    {...} & s>2\end{array}\right.\! \label{betas}.
    \end{equation}
The logic of the $\mbox{\boldmath$a$}$-theorem application is that the cosmological expansion can be associated with the transition from deep UV to IR regimes. The RG running in interacting and gravitating CFT can lead to the redistribution in the UV limit of the full set of conformal degrees of freedom to a higher spin domain, possessing according to the $\mbox{\boldmath$a$}$-theorem higher values of $\beta$ and $\tilde\beta$.  The present value of $\beta_{IR}=32\pi^2\mbox{\boldmath$a$}_{IR}$ can be a result of the evolution from a much larger initial value $\beta_{UV}=32\pi^2\mbox{\boldmath$a$}_{UV}$ responsible for the formation of a considerable thermal part of CMB.

\section{Conclusions}
New WMAP and Planck CMB data \cite{WMAP9} inspires interest in variety of modified vacuum and non-vacuum states of cosmological perturbations \cite{vacua}. The microcanonical state in the CFT cosmology has an advantage that it comes from first principles of quantum gravity \cite{slih,why} rather than from some ad hoc assumptions. Quite remarkably its formalism and physical predictions are determined by the Gauss-Bonnet anomaly and tightly related to the dilaton dynamics associated with the $\mbox{\boldmath$a$}$-theorem. In particular, stronger ``heating" of the CMB spectrum can be mediated by the RG flow interpolating between the UV and IR stages of cosmological expansion and, in view of this theorem, shifting the CFT model in UV to a higher spin phase. This opens a number of prospects for further research. Vacuum part of the spectrum determined by $|u_k(t)|^2$ in (\ref{occupation}) is needed: though only the inflaton mode remains dynamical, its effect for large $\beta$ is strongly mediated by the dilaton $\sigma$ whose sound cone in (\ref{kinetic}) might cause  superluminal phenomena \cite{superluminality}. The efficiency of higher-spin mechanism should be verified, as its RG increment $\mbox{\boldmath$a$}_{UV}-\mbox{\boldmath$a$}_{IR}$ is advocated to be always bounded \cite{PolchinskiRattazzi}. Moreover, {\em gravitating} higher-spin conformal fields do not seem to be explicitly known yet except $s=3/2$ and $s=2$, and they are anticipated to suffer from unitarity violation caused by higher derivatives \cite{Tseytlinconf,higherspin} (cf. $\beta_{3/2}<0$ in (\ref{betas}) for Weyl invariant gravitino with a third order wave operator \cite{Tseytlinconf}). Thus, the progress here strongly depends on advancing theory of conformal higher spin models \cite{higherspin}. There is a lot more to be learned within a remarkable interplay between $\mbox{\boldmath$a$}$-theorem and physics of the very early Universe.

{\bf Acknowledgements.} I am grateful to J.Garriga and S.Sibiryakov for helpful discussions and acknowledge support at the workshop YITP-T-12-03. This work was supported by the RFBR grant No. 11-02-00512.


\begin{thebibliography}{99}
\bibitem{KomargodskiSchwimmer}Z. Komargodski and A. Schwimmer, JHEP 1112 (2011) 099.

\bibitem{Komargodski}Z. Komargodski, JHEP 1207 (2012) 069.

\bibitem{Zamolodchikov}A. B. Zamolodchikov, JETP Lett. {\bf 43} (1986) 730.
    
\bibitem{matching}I. Jack and H. Osborn, Nucl. Phys. {\bf B343} (1990) 647; A. Schwimmer and S. Theisen, Nucl. Phys. {\bf B847} (2011) 590.

\bibitem{ChibisovMukhanov}
    V. F. Mukhanov and G. V. Chibisov, JETP Lett. {\bf 33} (1981) 532; V. Mukhanov, H. Feldman and R. Brandenberger, Phys. Rep. {\bf 215} (1992) 203.

\bibitem{slih}A. O. Barvinsky and A. Yu. Kamenshchik, JCAP {\bf 09} (2006) 014; Phys. Rev. {\bf D74} (2006) 121502.

\bibitem{why}A. O. Barvinsky, Phys. Rev. Lett. {\bf 99} (2007) 071301.

\bibitem{bigboost}
    A. O. Barvinsky, C. Deffayet and A. Yu. Kamenshchik, JCAP {\bf 05} (2008) 020; JCAP {\bf 05} (2010) 034.

\bibitem{anomalyaction}R.J.Riegert, Phys. Lett. {\bf B134} (1984) 56;
    E.S.Fradkin and A.A.Tseytlin, Phys. Lett. {\bf B134} (1984) 187;
    I.Antoniadis, P.O.Mazur and E.Mottola, Phys. Lett. {\bf  B323} (1994) 284;
    A.O.Barvinsky, A.G.Mirzabekian and V.V.Zhytnikov, {\em Conformal decomposition of the effective action and covariant curvature expansion}, gr-qc/9510037.

\bibitem{DGP}G.R.Dvali, G.Gabadadze and M.Porrati,  Phys. Lett.  {\bf B485} (2000) 208.

\bibitem{boxnablapi}M. A. Luty, M. Porrati and R. Rattazzi,
JHEP {\bf 0309} (2003) 029, arXiv:hep-th/0303116.

\bibitem{Vikmanetal}C.Deffayet, O.Pujolas, I.Sawicki and A.Vikman, JCAP {\bf 1010} (2010) 026.

\bibitem{CMBA-theorem}A. O. Barvinsky, {\em Thermal CMB in the CFT driven cosmology}, to be published.
    
\bibitem{ChristensenDuff}S.M.Christensen and M.J.Duff, Nucl. Phys. {\bf B154} (1979) 301.

\bibitem{Tseytlinconf}E.S.Fradkin and A.A.Tseytlin, Nucl. Phys. {\bf B203} (1982) 157; Phys. Rep. {\bf 119} (1985) 233.

\bibitem{WMAP9}G.Hinshaw et al, {\em Nine-Year WMAP Observations: Cosmological Parameter Results}, arXiv:1212.5226; P. A. R. Ade et al, {\em Planck 2013 Results.XXII. Constraints on Inflation}, arXiv:1303.5082.

\bibitem{vacua}I.Agullo and L.Parker, Phys. Rev. {\bf D83} (2011) 063526; J.Ganc and E.Komatsu, Phys.Rev.  {\bf D86} (2012) 023518; I. Agullo and S. Shandera, JCAP {\bf 1209} (2012) 007.

\bibitem{superluminality}A.Adams, N. Arkani-Hamed, S. Dubovsky, A. Nicolis and R. Rattazzi, JHEP {\bf 0610} (2006) 014.
    
\bibitem{PolchinskiRattazzi}  M. A. Luty, J. Polchinski, R. Rattazzi, JHEP {\bf 1301} (2013) 152.

\bibitem{higherspin}M.A.Vasiliev, Nucl. Phys. {\bf B829} (2010) 176; R. R. Metsaev, Mod. Phys. Lett. {\bf A10} (1995) 1719.
 



\end{thebibliography}
\end{document}